\documentclass[pra,twocolumn,showpacs,preprintnumbers,amsmath,amssymb]{revtex4}
\usepackage{mathrsfs}
\usepackage[hypertex,linkcolor=red]{hyperref}
\def\kk{\rangle\!\rangle}\def\bb{\langle\!\langle}\def\eg{e.~g. }

\def\rnk{\operatorname{rank}}\def\<{\langle}\def\>{\rangle}\def\Tr{\operatorname{Tr}}
\def\set#1{{\sf #1}}
\def\Rng{\set{Rng}}\def\Ker{\set{Ker}}\def\Supp{\set{Supp}}\def\Span{\set{Span}}
\def\dim{\operatorname{dim}}\def\sH{\set{H}}
\def\sE{\set{E}}\def\bP{{\mathbf P}}
\def\bD{{\mathbf D}}\def\bQ{{\mathbf Q}}
\def\conv#1{{\mathscr{#1}}}
\newtheorem{theorem}{Theorem}\newtheorem{corollary}{Corollary}
\begin{document}
\title{Classical randomness in quantum measurements}
\author{Giacomo Mauro D'Ariano}\email{dariano@unipv.it} 
\altaffiliation[Also at ]{Center for Photonic Communication and Computing, Department of
  Electrical and Computer Engineering, Northwestern University,
  Evanston, IL 60208}
\author{Paoloplacido Lo Presti}\email{lopresti@unipv.it} 
\author{Paolo Perinotti}\email{perinotti@fisicavolta.unipv.it} 
\altaffiliation[Also at ]{Istituto Nazionale di Fisica della Materia Unit\`a di Pavia}
\affiliation{{\em QUIT}  -- Quantum Information Theory Group,
\homepage{http://www.qubit.it} Dipartimento di Fisica
  ``A. Volta'', via Bassi 6, I-27100 Pavia, Italy} \date{\today}
\begin{abstract}
Similarly to quantum states, also quantum measurements can be "mixed", corresponding to a random
choice within an ensemble of measuring apparatuses. Such mixing is equivalent to a sort of
hidden variable, which produces a noise of purely classical nature. It is then natural to ask which
apparatuses are {\em indecomposable}, i. e. do not correspond to any random choice of apparatuses.
This problem is interesting not only for foundations, but also for applications, since most
optimization strategies give optimal apparatuses that are indecomposable.
\par Mathematically the problem is posed describing each measuring apparatus by a positive
operator-valued measure (POVM), which gives the statistics of the outcomes for any input
state. The POVM's form a convex set, and in this language the indecomposable apparatuses are
represented by extremal points---the analogous of "pure states" in the convex set of
states. Differently from the case of states, however, indecomposable POVM's are not necessarily 
rank-one, e. g. von Neumann measurements. 
\par In this paper we give a complete classification of indecomposable apparatuses
(for discrete spectrum), by providing different necessary and sufficient conditions for extremality
of POVM's, along with a simple general algorithm for the decomposition of  a POVM into extremals. 
As an interesting application, "informationally complete" measurements are analyzed in this
respect. The convex set of POVM's is fully characterized by determining its border in terms of
simple algebraic properties of  the corresponding POVM's. 
\end{abstract}
\pacs{03.65.Ta, 03.67.-a} \maketitle
\section{Introduction}
Measurements are the essence of any experimental science. At extreme sensitivities and precisions
they become the true core of quantum mechanics. For any practical need, a measurement can be 
always regarded as the retrieval of information about the "state" of the measured system. However, 
due to the no-cloning theorem\cite{Wootters,Yuen}, even for an elementary system---e.g. a single
harmonic oscillator or a spin---it is impossible to recover a complete knowledge of the state of
the system from a single measurement\cite{single} without prior knowledge. Then, since in 
quantum mechanics different incompatible measurements can be performed in principle, one is faced
with the problem of which measurement should be adopted for accomplishing a specific task, and 
which strategy of repeated measurements would be the most statistically efficient. These are the basic issues
of the operational viewpoint of Quantum Estimation Theory\cite{Helstrom}. 

A measurement on a quantum system \cite{Holevo} returns a
random result $e$ from a set of possible outcomes $\sE=\{e=1,\ldots
N\}$, with probability distribution $p(e|\rho)$ depending on the state
$\rho$ of the system in a way which is distinctive of the measuring
apparatus according to the Born rule
\begin{equation}
p(e|\rho)=\Tr[\rho P_e].\label{Born}
\end{equation}
In Eq. (\ref{Born}) $P_e$ denote positive operators on the Hilbert space $\sH$ of the system,
representing our knowledge of the measuring apparatus from which we infer information on the state
$\rho$ from the probability distribution $p(e|\rho)$. Positivity of $P_e$ is needed for positivity of
$p(e|\rho)$, whereas normalization is guaranteed by the completeness $\sum_{e\in\sE} P_e=I$. In the
present paper we will only consider the simple case of finite discrete set $\sE$. More generally, 
on has an infinite probability space $\set{E}$ (generally continuous), and in this context the set of 
positive operators $\{P_e\}$ becomes actually a positive operator valued measure (POVM). Every
apparatus is described by a POVM, and, reversely, every POVM can be realized in principle by an
apparatus \cite{bibnote,Nai,Ozawa,Holevo,Davies,Kraus,Busch}.
\par The linearity of the Born rule (\ref{Born}) in both arguments $\rho$ and $P_e$ is consistent
with the intrinsically statistical nature of the measurement, in which our partial knowledge of both
the system and the apparatus reflects in "convex" structures for both states and POVM's.  This
means that not only states, but also POVM's can be "mixed", namely there are POVM's that give 
probability distributions $p(e|\rho)$ that are equivalent to randomly choosing among different
apparatuses. Notice that mixed POVM's can also correspond to a single measuring apparatus, not only
when the apparatus itself is prepared in a mixed state, but also for pure preparation, as a result
of discarding (tracing out) the apparatus after a unitary interaction with the system. Clearly,
such mixing is itself a source of "classical" noise, which can be in principle removed by adopting an
indecomposable apparatus in the ensemble corresponding to the mixed POVM.
It is then natural to ask which apparatuses are {\em indecomposable}, 
i. e. "pure" in the above sense, or, mathematically, which POVM's correspond to extremal points of
the convex set. The classification of such apparatuses is certainly very useful in applications, since
most optimization strategies in quantum estimation theory\cite{Holevo,Helstrom} correspond to
minimize a concave (actually linear) function on the POVM's convex set---so-called
"cost-function"---whence leading to an optimal POVM which is extremal.  
\par Surprisingly, extremality for POVM's generally doesn't mean to be rank-one, as for "pure"
states. In other words, indecomposable POVM's are not necessarily realized by von Neumann 
measurements. Indeed, as we will see in this paper, there are rank-one POVM's that are not
extremal, whereas, on the opposite side,  there are higher-rank POVM's which are extremal. Moreover,
whenever the optimization problems have additional linear constraints---e. g. for covariant POVM's,
or for fixed probability distribution on a given state---the corresponding subset of POVM's is a
lower dimensional convex set corresponding to a section by a hyperplane of the complete POVM's convex set,
with boundary equal to the section of the original boundary, and whence with extremal points that
belong to the boundary of the convex set of all POVM's. For this reason, also the boundary of 
the POVM's convex set is interesting in practice, since POVM's that are optimal (for a concave cost
function) with an additional linear constraint generally are non extremal, but still 
belong to the boundary. One can also argue that POVM's which lie inside faces of the convex, 
physically exhibit a different degree of "classical" noise in relation with the dimensionality of
the face.  Notice that one should not imagine the POVM's set as a polytope, since, on the
contrary the set is ``strongly convex'', namely the extremal points are not isolated, but lie on
continuous manifolds.
\par In the present paper we address the problem of apparatus decomposability along three lines of
attack: a) by providing simple necessary and sufficient conditions for extremality of POVM's;
b) by establishing the complete structure of the POVM's convex set via the characterization of its
border in terms of algebraic properties of  the POVM's; c) by providing a simple
general algorithm for the decomposition of POVM's into extremals. For simplicity, the whole paper is
restricted to the case of discrete spectrum. In Section \ref{s:extr}, after clarifying the general features of
convex combinations of POVM's, using the method of perturbations we derive three different 
if-and-only-if conditions for extremality, along with some corollaries giving easy useful conditions,
only necessary or sufficient, that will be used in the following. Section \ref{s:qubits}
exemplifies the results of Section \ref{s:extr} in the case of a single qubit. Section \ref{s:boundary} 
presents the characterization of the border of the convex set in terms of algebraic
properties of POVM's. Section \ref{s:exmpl} shows that for every 
dimension $d=\dim(\sH)$ there is always an extremal POVM with maximal number $N$ of 
elements $N=d^2$, corresponding to a so-called informationally complete 
POVM\cite{infocompl}.  
After summarizing the results in the concluding Section \ref{s:concl}, Appendix \ref{a:decomp}
reports the algorithm for decomposing a POVM into extremals. 

\section{Convexity and extremality of POVM's}\label{s:extr}
Let's denote by ${\conv{P}_N}$ the convex set of POVM's on a finite dimensional
Hilbert space $\sH$, with a number $N$ of outcomes $\sE=\{1,\ldots,
N\}$. We will represent a POVM in the set as the vector $\bP\in{\conv{P}_N}$
$\bP=\{P_1,\ldots , P_N\}$ of the $N$ positive operators $P_e$. The fact
that the set ${\conv{P}_N}$ is convex means that it is closed under convex
linear combinations, namely for any $\bP',\bP''\in{\conv{P}_N}$ also
$\bP=p\bP'+(1-p)\bP''\in{\conv{P}_N}$ with $0\le p\le 1$ --- i.e. $p$ is a
probability. Then, $\bP$ can be also equivalently achieved by randomly
choosing between two different apparatuses corresponding to $\bP'$ and
$\bP''$, respectively, with probability $p$ and $1-p$, since, the
overall statistical distribution $p(e|\rho)$ will be the convex
combination of the statistics coming from $\bP'$ and $\bP''$. Notice
that ${\conv{P}_N}$ contains also the set of POVM's with a strictly smaller number of outcomes, i. e. with
$\sE'\subset\sE$. For such POVM's the elements corresponding
to outcomes in $\sE\setminus\sE'$ will be zero, corresponding to zero
probability of occurrence for all states. Clearly, for $N\le M$ one has ${\conv{P}_N}\subseteq\conv{P}_M
\subseteq\conv{P}$, where by $\conv{P}$ we denote the convex set of all POVM's with any
(generally infinite) discrete spectrum. The extremal points of ${\conv{P}_N}$ represent the "indecomposable"
measurements, which cannot be achieved by mixing different measurements. Obviously, a POVM which 
is extremal for ${\conv{P}_N}$ is also extremal for $\conv{P}_M$ with $M\ge N$, whence it is 
actually extremal for $\conv{P}$, and we will simply name it {\em extremal} without further
specifications.  
\par Let's start with a simple example. Consider the following two-outcome
POVM for a qubit
\begin{equation}
  {\mathbf P} = \left(\tfrac12|0\>\<0|,\,\tfrac12|0\>\<0|+|1\>\<1|\right).
\end{equation}
By defining $\bD=\frac12(|0\>\<0|,-|0\>\<0|)$, the two vectors
$\bP_\pm=\bP\pm\bD$ correspond to the following different POVM's
\begin{equation}
  {\mathbf P}_+ =\left(|0\>\<0|,\,|1\>\<1|\right)\,,\quad
  {\mathbf P}_- =\left(0,\,I\right),
\end{equation}
which, intuitively are extremal, whereas $\bP$ is not, since
$\bP=\frac12\bP_++\frac12\bP_-$ (we will see in the following that $\bP_\pm$ are indeed extremal).
\par Now the problem is how to assess when a POVM is extremal. Looking at the above
example, one notices that the non extremality of $\bP$ is equivalent to the existence of a vector
of operators $\bD\neq\mathbf{0}$ such that $\bP_\pm=\bP\pm\bD$ 
are POVM's, because in this case $\bP$ can be written as a convex
combination of $\bP_+$ and $\bP_-$. The non existence of such vector of operators
$\bD\neq\mathbf{0}$ is also a necessary condition for extremality of $\bP$, since for a non 
extremal $\bP$ there exist two POVM's $\bP_1\neq\bP_2$ such that
$\bP=\frac{1}{2}\bP_1+\frac{1}{2}\bP_2$, whence $\bD=\bP_1-\bP_2\neq\mathbf{0}$.
\par This leads to the method of perturbations for establishing 
extremality of a point in a convex set, which in the present context will be the following.
\medskip\paragraph*{Method of perturbations.}
We call a nonvanishing $\bD$ a {\em perturbation} for the POVM $\bP$ if there exists an 
$\epsilon>0$ such that $\bP\pm\epsilon\bD$ are both POVM's. Then a POVM is extremal if and only 
if it doesn't admit perturbations. 
\medskip
\par From the definition it follows that perturbations for POVM's are represented by vectors
$\bD$ of Hermitian operators $D_e$ (for positivity of $\bP\pm\epsilon\bD$) and
with zero sum $\sum_{e\in\set{E}}D_e=0$ (for normalization of $\bP\pm\epsilon\bD$). Specifically,
$\bD$ is a perturbation for $\bP$ if for some $\epsilon>0$ one has $P_e\pm\epsilon D_e\geq0$
$\forall e\in\set{E}$. Notice that the condition $P_e\pm\epsilon D_e\geq0$ is equivalent to  
$\epsilon|D_e|\leq P_e$, and a necessary condition for $\epsilon|D_e|\leq P_e$ is that 
$\Ker(P_e)\subseteq\Ker(D_e)$, or equivalently $\Supp(D_e)\subseteq\Supp(P_e)$
[the support $\Supp(X)$ of an operator $X$ is defined as the orthogonal complement of its 
kernel $\Ker(X)$]. In fact $\Ker(D_e)=\Ker(|D_e|)$, and for a vector $|\psi\>\in \Ker(P_e)$ with
$|\psi\>\not\in\Ker(|D_e|)$, one would have $\<\psi|(P_e-\epsilon|D_e|)|\psi\>=
-\epsilon\<\psi||D_e||\psi\>\le 0$, contradicting the hypothesis. 
Clearly, the Hermitian operators $D_e$ can be taken simply as linearly dependent---instead of having
zero sum---i.e.  $\sum_e\lambda_eD_e=0$ for non-vanishing $\lambda_e$, and, moreover, one can
consider more generally complex operators $D_e$ with 
$\Supp(D_e)\cup\Rng(D_e)\subseteq\Supp(P_e)$ $\forall e\in\set{E}$, and satisfying 
$\epsilon|D_e|\leq P_e$. In fact $P_e\pm\epsilon D_e'\geq0$ is satisfied $\forall e\in\set{E}$ by
the set of  Hermitian operators $D_e'=\lambda_eD_e+\lambda^*_eD_e^\dag$, for which $\sum_eD_e'=0$. 
\par The above considerations show that what really matters in assessing
the extremality of the POVM $\bP=\{P_e\}$ is just a condition on the supports $\Supp(P_e)$,
corresponding to the following theorem.
\begin{theorem}\label{t:lin_homog}
  The extremality of the POVM $\bP$ is equivalent to the nonexistence
  of non trivial solutions $\bD$ of the equation
\begin{equation}
\sum_{e\in\set{E}} D_e=0, \quad\Supp(D_e)\cup\Rng(D_e)\subseteq\Supp(P_e)
\;\forall e\in\set{E}.
\label{lin_homog}
\end{equation}
\end{theorem}
The above condition can be made explicit on the $P_e$ eigenvectors $\{|v^{(e)}_n\>\}$ corresponding
to nonzero eigenvalue, which therefore span $\Supp(P_e)$. Then, Eq. (\ref{lin_homog}) becomes the
linear homogeneous system of equations in the variables $D_{nm}^{(e)}=\< v^{(e)}_n|D_e|v^{(e)}_m\>$
\begin{equation}
\begin{split}
  \sum_{e\in\set{E}}
  \sum_{nm=1}^{\rnk(P_e)}&D_{nm}^{(e)}|v_n^{(e)}\>\<v_m^{(e)}|=0\\ 
&\Longleftrightarrow
D_{nm}^{(e)}=0\;\forall n,m,\,\forall e\in\set{E},
\label{eq:lin_co}
\end{split}
\end{equation}
namely the following version of Theorem \ref{t:lin_homog}
\begin{theorem}\label{c:partha}
A POVM $\bP=\{P_e\}_{e\in\set{E}}$ is extremal iff the operators $|v^{(e)}_n\>\<v^{(e)}_m|$ 
made with the eigenvectors of $P_e$ are linearly independent $\forall e\in\set{E}$ and 
$\forall n,m=1,\ldots,\rnk(P_e)$.
\end{theorem}
Theorem \ref{c:partha} is the characterization of extremal POVM's given by Parthasaraty in
Ref. \cite{Parthasaraty} in a $C^*$-algebraic setting. Notice that instead of the eigenvectors
$\{|v^{(e)}_n\>\}$ one can more generally consider a non orthonormal basis, which can be useful in
numerical algorithms.
\medskip
\par Another interesting way to state Theorem \ref{t:lin_homog} is in terms
of {\em weak independence} of  orthogonal projectors $Z_e$ on $\Supp(P_e)$,
where we call a generic set of orthogonal projections
$\{Z_e\}_{e\in\set{E}}$ weakly independent \cite{stormer} if for any set of operators 
$\{T_e\}_{e\in\set{E}}$ on $\sH$
one has 
\begin{equation}
\sum_{e\in\set{E}}Z_e T_eZ_e=0\quad\Rightarrow\quad Z_e T_e Z_e=0,\;\forall e\in\set{E}.
\end{equation}
Notice that since the orthogonal projectors are in one-to-one correspondence with their supporting
spaces, the notion of weak independence can be equivalently attached to the supporting spaces
$\Supp(P_e)$. This leads us to the alternative characterization of extremal POVM's
\begin{theorem}\label{t:weak}
A POVM $\bP=\{P_e\}_{e\in\set{E}}$ is extremal iff the supports $\Supp(P_e)$ are weakly 
independent for all $e\in\set{E}$. 
\end{theorem}
The proof is straightforward if one considers that any perturbation
$\bD$ for $\bP$ can be written as $D_e=Z_eD_eZ_e$ with the constraint
\begin{equation}
\sum_eZ_eD_eZ_e=0\,.
\end{equation}
Then the theorem simply says that the only allowed perturbation for an extremal POVM is the trivial 
one.
\medskip
\par  Some corollaries relevant for applications follow immediately from the main Theorem
\ref{t:lin_homog} or its equivalent versions. 
\begin{corollary}\label{c:rank}
For $\sum_{e\in\set{E}}\dim[\Supp (P_e)]^2>d^2$ the POVM $\bP$ is not extremal.
\end{corollary}
This means that a POVM with more than $d^2$ nonvanishing elements is always decomposable into 
POVM's with less than $d^2$ elements. For the case of $d^2$ elements Theorem \ref{t:lin_homog} 
also implies that
\begin{corollary}\label{c:d2}
An extremal POVM with $d^2$ outcomes must be necessarily rank-one.
\end{corollary}
In fact, for larger rank we would have more than $d^2$ eigenvectors $|v_n^{(e)}\>$, 
and the operators $|v^{(e)}_n\>\<v^{(e)}_m|$ cannot be linearly independent. From Theorem
\ref{t:lin_homog} it also follows that if some elements of the POVM $\bP$ have non-disjoint
supports, then $\bP$ is not extremal, or, equivalently 
\begin{corollary}\label{s:disj}
An extremal POVM $\bP=\{P_e\}$ must have all supports $\Supp(P_e)$ mutually disjoint.
\end{corollary}
Precisely, we call two linear spaces $\set{A}$ and $\set{B}$ {\em disjoint} when 
$\set{A}\cap\set{B}=\{0\}$ is the null vector. It is worth noticing that two linear spaces that are
disjoint are not necessarily orthogonal, whereas, reversely, two orthogonal spaces are clearly
disjoint. We emphasize that the condition of Corollary \ref{s:disj} is only necessary. Indeed, it is
easy to envisage a POVM that satisfies the above condition without being extremal, e. g. a rank-one
POVM for $d=2$ with five elements corresponding to the vertices of a pentagon in the Bloch sphere
(see Section \ref{s:qubits} for extremal POVM's for qubits).
\par Other obvious consequences of Theorem \ref{c:partha} are the following
\begin{corollary}
Orthogonal POVM's are extremal.
\end{corollary}
\begin{corollary}\label{c:rnk1}
  A rank-one POVM is extremal if and only if its elements $P_e$ are
  linearly independent.
\end{corollary}
Corollary \ref{c:rank} states that for dimension $d$ an extremal POVM can have at most $d^2$ 
non-null elements (in Section \ref{s:exmpl} we will show that such an extremal POVM always exists). 
Here we can immediately conclude that
\begin{corollary}\label{c:infocom}
A POVM with $d^2$ elements is necessarily a rank-one "informationally complete" POVM.
\end{corollary}
By definition, an informationally complete POVM $\bP=\{P_e\}$ \cite{infocompl} has elements $P_e$ 
which span the space of all operators on $\sH$, thus allowing the estimation of any ensemble average
using the same fixed apparatus. The fact that a POVM with $d^2$ elements necessarily is rank-one is
stated in Corollary \ref{c:d2}, whereas Corollary \ref{c:rnk1} assures that all POVM's elements are
linearly independent, whence the set of $d^2$ linearly independent operators $P_e$ is obviously
complete for dimension $d$. In Section \ref{s:exmpl} we will give an explicit example
of such an extremal informationally complete POVM.
\par For rank greater than one we have only the necessary condition
\begin{corollary}\label{c:indep}
  If the POVM is extremal, then its non-vanishing elements are
  necessarily linearly independent.
\end{corollary}
In fact, according to Theorem \ref{c:partha} the projectors on the eigenvectors must be linearly 
independent, whence also the operators $P_e$. Indeed, for linearly dependent elements there exist
coefficients $\lambda_e$ not all vanishing such that $\sum_{e\in\set{E}}\lambda_e P_e=0$, and
without loss of 
generality we can take $-1\le \lambda_e\le 1$. Therefore, the POVM can be written as convex
combination  $\bP=\frac{1}{2}\bP^-+\frac{1}{2}\bP^+$, with $P_e^\pm=(1\pm\lambda_e)P_e$,
and $\bP^-\neq\bP^+$ (since  the $\lambda_e$'s are not all vanishing). 
\par In appendix \ref{a:decomp} we report an algorithm for decomposing a given POVM into extremal. 
\section{Extremal POVM's for qubits}\label{s:qubits}
Using the above results we will give a classification of extremal POVM's for qubits. In this case,
Corollary \ref{c:rank} implies that the extremal POVM's cannot have more than 4 elements, and that,
apart from the trivial POVM $\bP=I$, they must be made of rank-one projectors (otherwise 
$\Supp(P_e)$ for different $e\in\set{E}$ would not be mutually disjoint). Now, upon writing the 
POVM elements in the Bloch form 
\begin{equation}
P_e=\alpha_e\left(I+\vec n_e\cdot\vec\sigma\right)\,.
\label{eq:povmqubit}
\end{equation}
the constraints for normalization and positivity read
\begin{equation}
\alpha_e>0\,,\quad\sum_e\alpha_e=1\,,\quad\sum_e\alpha_e\vec n_e=0\,.
\label{eq:constraints}
\end{equation}
The case of two outcomes corresponds simply to the usual observable 
$P_e=|e\>_{\vec n}\;{}_{\vec n}\<e|$, $e=0,1$, with $|e\>_{\vec n}$ eigenvector of
$\vec n\cdot\vec\sigma$ corresponding to the eigenvalues $+1, -1$, respectively. In fact, for two
outcomes one has $\alpha_0\vec n_0+\alpha_1\vec n_1=0$, namely $\vec n_0=-\vec
n_1\doteq\vec n$, and necessarily $\alpha_0=\alpha_1=\frac12$.
We now consider the case of 3 and 4 elements. By Theorem \ref{c:partha} a
necessary and sufficient condition for extremality is 
\begin{equation}
\sum_{e\in\set{E}}\gamma_e\alpha_e(I+\vec n_e\cdot\vec\sigma)=0
\Longleftrightarrow \gamma_e=0\;\forall e\in\set{E},
\label{eq:lincomb1}
\end{equation}
or, equivalently,
\begin{equation}
\sum_{e\in\set{E}}\gamma_e\alpha_e=0,\;\sum_{e\in\set{E}}\gamma_e\alpha_e\vec n_e=0
\Longleftrightarrow\gamma_e=0\;\forall e\in\set{E}.
\label{eq:lincomb2}
\end{equation}
For three outcomes Eq.  (\ref{eq:constraints}) implies that
$\{\alpha_e\vec n_e\}_{e\in\set{E}}$ represent the edges of a triangle, and thus the
second condition in Eq. (\ref{eq:lincomb2}) is satisfied iff
$\gamma_e\equiv\gamma$ is independent of $e$. Then the first condition
is satisfied iff $\gamma\equiv0$. Therefore, all three outcomes rank-one POVM's 
with pairwise non proportional elements are extremal. For four outcomes
we can see that for an extremal POVM the corresponding unit vectors $\{\vec n_e\}_{e\in\set{E}}$
cannot lie on a common plane. Indeed, divide the four vectors 
$\{\alpha_e\vec n_e\}_{e\in\set{E}}$ into two couples, which identify two intersecting planes. Then,
the third condition in Eq. (\ref{eq:constraints}) implies that the sums of the
 couples lie on the intersection of the planes and have the same
length and opposite direction. If we multiply by independent scalars
$\gamma_e$ the two elements of a couple, their sum changes direction
and lies no more in the intersection of the two planes, and the second
condition in Eq. (\ref{eq:lincomb2}) cannot be satisfied. Therefore, the two
elements of the same couple must be multiplied by the same
scalar, which then just rescales their sum. Now, when the rescaling
factors are different for the two couples the two partial sums don't
sum to the null vector anymore. Then necessarily 
$\gamma_e\equiv\gamma$ independently of $e$. In order to satisfy also
the first condition in Eq. (\ref{eq:lincomb2}) we must have
$\gamma=0$. On the other hand, if the four unit vectors lie in the same
plane, a non-trivial linear combination can always be found that
equals the null operator, hence the POVM is not extremal. By the first
condition in Eq. (\ref{eq:lincomb2}) we have indeed
\begin{equation}
\gamma_0\alpha_0=-\sum_{e=1,2,3}\gamma_e\alpha_e\,,
\end{equation}
then the second condition can be written
\begin{equation}
\sum_{e=1,2,3}\gamma_e\alpha_e(\vec n_e-\vec n_0)=0\,.
\label{eq:lincomb3}
\end{equation}
Now, either we have a couple of equal vectors $\vec n_e$, or the three
vectors $\vec n_e-\vec n_0$ in a two dimensional plane are linearly
dependent. However, in both cases the POVM is not extremal, because in
the former case two elements are proportional, while in the latter a
non trivial triple of coefficients $\gamma_e$ satisfying Eq.
(\ref{eq:lincomb3}) exists.\par

Notice that, the three and four-outcomes POVM's are necessarily unsharp, i.e. there is no state
with probability distribution $p(e|\rho)=\delta_{e,\bar e}$ for a fixed $\bar e$. They provide 
examples of un-sharp POVM's with purely intrinsical quantum noise.

\section{The boundary of the convex set of POVM's}\label{s:boundary}

In this section we generalize the results about extremality, and
give a full characterization of the elements on the boundary of the
convex set of $n$-outcomes POVM's on the Hilbert space $\sH$. 
Let start from an intuitive geometrical definition of the boundary of a convex set. Consider for
example a point lying on some face of a polyhedron. Then there exists a
direction (e. g. normal to the face) such that any shift of the point along
that direction will bring it inside the convex set, while in the opposite direction it will bring
the point outside of the convex. In mathematical terms, consider a convex set $\conv{C}$ and 
an element $p\in\conv{C}$. Then, $p$ belongs to the boundary $\partial\conv{C}$ of $\conv{C}$
if and only if there exists $q\in\conv{C}$ such that
\begin{equation}
p+\epsilon(q-p)\in\conv{C}\,,\quad
p-\epsilon(q-p)\not\in\conv{C}\,,\quad\forall\epsilon\in[0,1]. 
\end{equation}
This definition leads to the following characterization of the boundary of the convex set of 
$N$-outcomes POVM's
\begin{theorem}\label{t:boundary}
  A POVM $\bP\in{\conv{P}_N}$ belongs to the boundary of ${\conv{P}_N}$ iff at least one
  element $P_f$ of $\bP$ has a non trivial kernel.
\end{theorem}
Let's first prove necessity. Consider two different POVM's $\bP$ and
$\bQ$, and suppose that $\forall\epsilon\in[0,1]$ $\bP+\epsilon \bD$ is still a POVM while
$\bP-\epsilon \bD$ is not.  This happens only if $\forall\epsilon\in[0,1]$ $P_f-\epsilon
D_f\not\geq0$ for some $f$. Then some vector $\psi$ must exist such that
\begin{equation}
\<\psi|P_f|\psi\><\epsilon\<\psi|D_f|\psi\>,\quad\forall\epsilon\in[0,1],
\end{equation}
namely $\<\psi|P_f|\psi\>=0$. Since by hypothesis $P_f$ is positive
semidefinite, then necessarily $\psi\in\Ker(P_f)$. To prove that the
condition is also sufficient, consider a POVM element $P_f$ with nontrivial kernel, and take
$\psi\in\Ker(P_f)$. Then consider an event $g$ such that $\<\psi|P_g|\psi\>>0$ 
(such event must exist for normalization of the POVM), and take $D_f=\kappa|\psi\>\<\psi|$,
$D_g=-\kappa|\psi\>\<\psi|$, and $D_e=0$ otherwise, with $\kappa$ smaller than the
minimum eigenvalue of $P_g$. Clearly $\forall\epsilon\in[0,1]$ $\bP+\epsilon \bD$ is a
POVM while $\bP-\epsilon\bD$ is not, since the element $P_f-\epsilon
D_f$ is not positive semidefinite.
\medskip
\par We now proceed to study the structure of the faces of ${\conv{P}_N}$. For such purpose it is convenient to
regard a convex set as a subset of an affine space, whose dimension is the number of linearly
independent directions along which any internal point can be symmetrically shifted. Clearly, also
the faces of the convex set are themselves convex. For example, moving from a point inside of a cube
one can explore three dimensions while remaining inside, whereas, for the cube faces the
number of independent symmetric perturbations is two, and for the sides this number reduces to
one. We will keep in mind the above geometrical picture for the classification of the border of the
convex set of POVM's using the perturbation method.
\par According to the results of Section \ref{s:extr}, a perturbation for a POVM $\bP$ is a set
of Hermitian operators $\bD=\{D_e\}$ with $\sum_{e\in\set{E}}D_e=0$, and with
$\Supp(D_e)\subseteq\Supp(P_e)$. Expressed in the orthonormal basis of eigenvectors of 
the POVM elements as in Eq. (\ref{eq:lin_co}), the operators $D_e$ read
\begin{equation}
D_e=\sum_{mn=1}^{\rnk(P_e)}D^{(e)}_{mn}|v^{(e)}_m\>\<v^{(e)}_n|.
\end{equation}
We remind that, according to Theorem \ref{c:partha}, non trivial perturbations for $\bP$ exist only if
the outer products $|v^{(e)}_m\>\<v^{(e)}_n|$ are linearly dependent with $e\in\set{E}$ and 
$1\le n,m\le\rnk(P_e)$. The total number of
such outer products is
\begin{equation}
r(\bP)\doteq\sum_{e\in\set{E}}\rnk(P_e)^2.\label{rP}
\end{equation}
The number of linearly independent elements in the set of outer products 
$|v^{(e)}_m\>\<v^{(e)}_n|$ is given by
\begin{equation}
l(\bP)\doteq\dim[\Span(\{|v^{(e)}_m\>\<v^{(e)}_n|\})].
\end{equation}
Now, as in Theorem \ref{t:boundary}, we see that the number $b(\bP)$ of independent
perturbations for $\bP$ is given by
\begin{equation}
b(\bP)=r(\bP)-l(\bP).\label{bP}
\end{equation}
In fact, in Eq. (\ref{eq:lin_co}) we have $r(\bP)=\sum_{e\in\set{E}}\rnk(P_e)^2$ variables
$D_{nm}^{(e)}$, whereas the number of linearly independent equations is
$l(\bP)=\dim(\Span(\{|v^{(e)}_m\>\<v^{(e)}_n|\}))$, whence the number of variables which 
can be written as linear combination of a linearly independent set is $r(\bP)-l(\bP)$.
On the other hand, the dimension of the affine space of the convex set
${\conv{P}_N}$ is given by $d^2(N-1)$, since the POVM normalization constraint corresponds to $d^2$ 
independent linear equations, with $Nd^2$ variables. We have then proved the following
characterization of the border of ${\conv{P}_N}$
\begin{theorem}\label{t:boundirs}
  A POVM $\bP\in{\conv{P}_N}$ belongs to the boundary $\partial\conv{P}_N$ of ${\conv{P}_N}$ iff
  $b(\bP)<d^2(N-1)$, $b(\bP)$ defined in Eqs. (\ref{rP}-\ref{bP}) being the dimension of the face in
  which $\bP$ lies. 
\end{theorem}
From the above theorem it also follows that a POVM $\bP\in{\conv{P}_N}$ on the boundary
$\partial\conv{P}_N$ of ${\conv{P}_N}$ also belongs to $\partial\conv{P}_M$ with $M\ge N$, whence it
belongs to the boundary $\partial\conv{P}$ of $\conv{P}$. This also implies that
$\partial\conv{P}_N\subseteq\partial\conv{P}_M \subseteq\partial\conv{P}$. 
\section{Extremal informationally complete POVM's}\label{s:exmpl}
In this section we will give an explicit construction of a rank-one informationally complete POVM as
in Corollary \ref{c:infocom}, in this way also proving the existence of extremal POVM's with $d^2$
elements. 
\par Consider the {\em shift-and-multiply} finite group of unitary operators 
\begin{equation}
U_{pq}=Z^pW^q,\quad p,q\in{\mathbb Z}_d
\end{equation}
where ${\mathbb Z}_d=\{0,1,\ldots,d-1\}$, and $Z$ and $W$ are defined as follows
\begin{equation}
Z=\sum_j|j\oplus1\>\<j|,\; W=\sum_j\omega^j|j\>\< j|,
\end{equation}
with $\oplus$ denoting the sum modulo $d$, $\{|j\>\}_0^{d-1}$ an orthonormal basis,
$\omega=e^{\frac{2\pi i}d}$, and the sums are extended to ${\mathbb Z}_d$. We now prove that the
following POVM with $d^2$ outcomes is  
extremal 
\begin{equation}
P_{pq}\doteq \frac1d U_{pq}\nu U_{pq}^\dag\,,
\label{eq:d2extr}
\end{equation}
for any pure state $\nu=|\psi\>\<\psi|$ on $\sH$ satisfying the constraints
\begin{equation}
\Tr[U_{pq}^\dag\nu]\neq 0,\quad \forall p,q\in{\mathbb Z}_d.\label{constrnu}
\end{equation}
In order to prove the statement, first we notice that the operators $d^{-\frac{1}{2}}U_{pq}$ form
a complete orthonormal set of unitary operators, i. e. they satisfy 
\begin{eqnarray}
\Tr[U_{pq}U_{p',q'}]&=&d\delta_{pp'}\delta_{qq'},\\\sum_{pq}U_{pq}\Xi
U_{pq}^\dag&=&d\Tr[\Xi],\label{trcU}
\end{eqnarray}
for any operator $\Xi$.
From Eq. (\ref{trcU}) it immediately follows that $\sum_{pq}P_{pq}=I$, whence $\{P_{pq}\}$ is a 
POVM. Then, in order to prove extremality, according to Corollary \ref{c:rnk1} it is sufficient to
prove that the $d^2$ operators $P_{pq}$ are linearly independent, which in turn can be proved by
showing that $\{P_{pq}\}$ is itself a complete set in the space of operators (completeness along
with the fact that the elements $P_{pq}$ are $d^2$ implies indeed that they are linearly
independent). As mentioned after Corollary \ref{c:infocom}, such a kind of completeness for the POVM 
corresponds to a so-called {\em informationally complete measurement}\cite{infocompl}.
The completeness of the set $\{P_{pq}\}$ is equivalent to the
invertibility of the following operator on $\sH^{\otimes 2}$
\begin{equation}
F=\sum_{pq}|P_{pq}\kk\bb P_{pq}|,\label{Fop}
\end{equation}
where the double-ket notation\cite{bellobs} 
is used to remind the equivalence between (Hilbert-Schmidt) operators $A$ on
$\sH$ and vectors $|A\kk=A\otimes I|I\kk$ of $\sH^{\otimes 2}$, $|I\kk\in\sH^{\otimes 2}$ 
denoting the reference vector $|I\kk=\sum_{j\in{\mathbb Z}_d}|j\>\otimes|j\>$ defined in terms of
the chosen orthonormal basis $\{|j\>\}$. By expanding $\nu=|\psi\>\<\psi|$ over the basis
$\{U_{pq}\}$ and using the multiplication rules of the group, one obtains
\begin{equation}
P_{pq}=\frac{1}{d^2}\sum_{rs} e^{\frac{2\pi i}{d}(qr-ps)}
\Tr[U_{rs}^\dag\nu]U_{rs},  
\end{equation}
which allows us to rewrite Eq. (\ref{Fop}) as follows
\begin{equation}
F=\frac{1}{d^2}\sum_{rs}|\Tr[U_{rs}^\dag\nu]|^2|U_{rs}\kk\bb U_{rs}|.
\end{equation}
Since $\{U_{pq}\kk\}$ is an orthogonal basis in $\sH^{\otimes 2}$, the invertibility of $F$ is
equivalent to condition (\ref{constrnu}), which is clearly satisfied by most density operators
$\nu=|\psi\>\<\psi|$ (condition (\ref{constrnu}) is satisfied by a set of states $|\psi\>$ that is
dense in $\sH$). As an example of 
state satisfying the condition (\ref{constrnu}), one can consider 
$|\psi\>\propto\sum_j\alpha^j|j\>$ for $0<\alpha<1$. 
\section{Conclusions}\label{s:concl}
In this paper we have completely characterized the convex set of POVM's with discrete
spectrum. Using the method of perturbations, we have determined the extremal
points---the "indecomposable apparatuses"---by three alternative characterizations corresponding to
Theorems  \ref{t:lin_homog}-\ref{t:weak}, and with easier necessary or sufficient conditions in
Corollaries \ref{c:rank}-\ref{c:indep}. In particular, we have shown that for finite dimension $d$
an extremal POVM's can have at most $d^2$ outcomes, and an extremal POVM with $d^2$ outcomes
always exists and is necessarily informationally complete. An explicit realization of such extremal
informationally complete POVM has been given in Section \ref{s:qubits}.
\par The characterization of the convex set ${\conv{P}_N}$ of POVM's with $N$ outcomes has been obtained by
determining its boundary $\partial\conv{P}_N$,  which, in turn, has been characterized in terms of the
number $b(\bP)$ of independent perturbations for the POVM $\bP$ in Eq. (\ref{bP}). This has lead to
a simple characterization of the boundary in terms of simple
algebraic properties of a POVM lying on it. Since $\partial\conv{P}_N$ is also a subset of the boundary 
$\partial\conv{P}$ of the full convex set $\conv{P}$ of POVM's with discrete spectrum, our result also
provides a complete characterization of $\conv{P}$. 
\par Finally, in Appendix \ref{a:decomp} we reported an algorithm for decomposing a point in a
convex set into a minimum number of extremal elements, specializing the algorithm to the case of the
convex set ${\conv{P}_N}$ of POVM's with $N$ outcomes.
\appendix
\section{Algorithm for decomposition of internal points into extremals}\label{a:decomp}
In this appendix we provide an algorithm to decompose a POVM $\bP\in{\conv{P}_N}$ into extremal 
ones. We first present the algorithm in the general case of an abstract convex set, and then we
specialize it to the case of POVM's.
\smallskip
\par Consider a convex set $\conv{C}$ and a point $p$ inside it. We want to decompose $p$ into extremal
points. We first need two ingredients, which depend on the specific convex set $\conv{C}$ under
consideration: the {\em affine space} $\conv{A}$ in which $\conv{C}$ is embedded (this is just the space of
legitimate "perturbations"), and an "indicator" $\iota(p)$ which is positive for $p\in\conv{C}$, zero
on the boundary, and negative outside $\conv{C}$. Now, starting from $p$ inside $\conv{C}$, 
we move in some direction $d$ in $\conv{A}$ until a face of $\conv{C}$ is encountered at
$\lambda_+\doteq\max\{\lambda:p+\lambda d\in\conv{C}\}$ 
(using our indicator $\lambda_+$ is 
given by the value of $\lambda$ where $\iota(p+\lambda d)$ changes sign).
Similarly in the opposite direction one hits the boundary at 
$\lambda_-\doteq\max\{\lambda:p-\lambda d\in\conv{C}\}$. 
The point $p$ can now be split into the convex combination
\begin{equation}
p=\frac{\lambda_-}{\lambda_++\lambda_-}p_++
\frac{\lambda_+}{\lambda_++\lambda_-}p_-,\quad p_\pm\doteq p\pm\lambda_\pm d.
\end{equation}
If $p$ was in the interior of $\conv{C}$ any face of the boundary can be encountered, while if $p$ was
already on the boundary of $\conv{C}$ the perturbation $d$ brings $p$ on a "face of a face'', \eg moving on a
face of a cube toward an edge. In any case, the dimension $b(p)$ of the face to which $p$ belongs is
decreased at least by one.
\par By applying the same splitting scheme to both $p_\pm$ recursively,
we obtain a weighted binary "tree" of points rooted in $p$, with the
property that the point $p'$ at each node can be written as convex
combination of its descendants, and with a depth bounded by the
dimension $b(p)$ of the face to which $p$ belongs. Of course the "leaves" of the tree are extremal
points $p_i$, and one can combine them to obtain the original point as $p=\sum_i\alpha_ip_i$
weighting each leaf $p_i$ with the product $\alpha_i$ of all weights found along the path from the
root $p$ to the leaf $p_i$. Unfortunately, this raw algorithm can produce up to $2^r$
extremal points $p_i$, for dimension $r$ of the affine space of $\conv{C}$, each leaf being
addressed by the vector $d_i=p-p_i$, with $\sum_i\alpha_id_i=0$. However, by the Caratheodory's 
theorem \cite{Rockafellar}, we know that at most $r+1$ extremal point are needed to decompose 
$p$. Indeed, if the number of $d_i$ is larger than $r+1$, then they must be linearly dependent, and
there must exist $\lambda_i$'s not all vanishing and not all positive such that
$\sum_i\lambda_id_i=0$. Since 
$\sum_i\alpha_id_i-\mu(\sum_i\lambda_id_i)=0$, by choosing the
greatest $\mu$ such that $\alpha_i-\mu\lambda_i\geq 0$ $\forall i$,
one finds that $0$ can be written as a convex combination of a
smaller number of $d_i$'s. This procedure can be applied repeatedly
to the remaining $d_i$'s (the $\alpha_i$'s must also be upgraded) until
their only combination giving $0$ is the one whose coefficients are all positive: at this point
their number is for sure not larger than $r+1$.  Therefore, from an intial decomposition with
many elements, we end up with a decomposition of $p$ into at most $r+1$ vectors $d_i$ and
probabilities $\alpha_i$ such that $\sum_{i=1}^{r+1}\alpha_id_i=0$, whence
$p=\sum_{i=1}^{r+1}\alpha_i(p+d_i)=\sum_{i=1}^{r+1}\alpha_ip_i$.
Notice that the evaluation of $\lambda_\pm$ at each step involves an
eigenvalue evaluation, whence the algorithm generally doesn't provide
analytical decompositions. 
\medskip
\par In order to specialize the algorithm to the case of the POVM's convex set $\conv{P}_N$, we
need to specify both the corresponding affine space $\conv{A}_N$ and the indicator $\iota$ of the
border. The affine space is the real $d^2(N-1)$-dimensional linear space of  vectors $\bD=\{D_e\}$ of
$N$ Hermitian operators $D_e$, with $\sum_{e\in\set{E}}D_e=0$. This can be obtained as the real 
span of  projectors  
$|n,e\>\<n,e|$ along with $\Re|n,e\>\< m,e|$  and $\Im|n,e\>\< m,e|$ for $n=1,\ldots,d$ 
and $n<m$, where $X=\Re X+i\Im X$ is the Cartesian decomposition of the operator $X$, 
$\{|n,e\>\}$ denotes any orthonormal basis for the $e$-th copy of the Hilbert 
space $\sH$ of the quantum system with $d=\dim(\sH)$, and $e\in\set{E}'\doteq\set{E}/\{1\}$, and
$D_1=-\sum_{e\in\set{E}'}\operatorname{diag}D_e$. However, if such global basis is used, then 
when the search algorithm starts from a POVM which is already on a face of the convex one
has the problem that generally the basis is not aligned with the face itself, and for a generic
direction the perturbed POVM either exits from the convex, or it moves inside it. For that reason it is
convenient to consider a "local basis" of perturbations for a given $\bP$. This can be
constructed by considering the set $\{X^{(e)}_{m,n}\}$ of Hermitian operators defined as follows 
\begin{equation}
\begin{split}
X^{(e)}_{mn}&=\Re|v^{(e)}_m\>\<v^{(e)}_n|,\quad
X^{(e)}_{nm}=\Im|v^{(e)}_m\>\<v^{(e)}_n|,\quad n< m\\
X^{(e)}_{nn}&=|v^{(e)}_n\>\<v^{(e)}_n|.
\end{split}
\end{equation}
Then pick up $l(\bP)$ linearly independent elements, which we will denote by
$V^{(e)}_{mn}$, and call the remaining ones $W^{(f)}_{mn}$, so that we can write
\begin{equation}\label{Ws}
W^{(f)}_{mn}=\sum_{epq}c^{fmn}_{epq}V^{(e)}_{pq}\equiv
\sum_{e\neq f}\sum_{pq}c^{fmn}_{epq}V^{(e)}_{pq},
\end{equation}
where the second identity is a consequence of linear independence of operators
$\{|v^{(e)}_m\>\<v^{(e)}_n|\}$ for fixed $e$. Then we can construct the following basis
$\{D_e(fmn)\}$ for $\bP$ perturbations 
\begin{equation}
\begin{split}
&D_f(fmn)=\sum_{pq}c^{fmn}_{fpq}V^{(f)}_{pq}-W^{(f)}_{mn},\\
&D_e(fmn)=\sum_{pq}c^{fmn}_{epq}V^{(e)}_{pq}.\label{eq:pertu}
\end{split}
\end{equation}
Clearly one has $\sum_{e\in\set{S}}D_e(fmn)=0$ $\forall fmn$, and modulo a suitable rescaling one
has $P_e\pm D_e(fmn)\geq0$ $\forall fmn$. Notice that, by construction, the operators $D_e(fnm)$ are
linearly independent. In fact, using Eqs. (\ref{eq:pertu}), a generic linear combination of
$D_e(fmn)$ for each fixed $e\in\set{E}$ will result in a linear combination 
of $\{V^{(e)}_{mn}\}\cup\{W^{(e)}_{pq}\}$ for the same $e$. But the set
$\{V^{(e)}_{mn}\}\cup\{W^{(e)}_{pq}\}$ is linearly independent for fixed $e\in\set{E}$, due to 
Eq. (\ref{eq:pertu}). 
\par As regards the indicator of the boundary, this is simply the minimum of the eigenvalues of  the
operators $P_e$, which changes from positive to negative when $\bP$ crosses the border. 
\acknowledgements
We are grateful to R. Mecozzi for his participation at the preliminary stage of this research during his
thesis work. This work has been co-founded by the EC under the program ATESIT (Contract No.
IST-2000-29681) and the MIUR cofinanziamento 2002. P.P. acknowledges support from the INFM under
project PRA-2002-CLON. 
 
\end{document}